\documentclass[twocolumn,showpacs,preprintnumbers,amsmath,amssymb,prl,superscriptaddress]{revtex4-1}
\usepackage{amsfonts}
\usepackage[english]{babel}
\usepackage[T1]{fontenc}
\usepackage{times}
\usepackage{mathrsfs}
\usepackage{graphicx}
\usepackage{dcolumn}
\usepackage{bm}
\usepackage[colorlinks,bookmarks=true,citecolor=blue,linkcolor=red,urlcolor=blue]{hyperref}
\usepackage[tight, FIGTOPCAP, hang, raggedright, nooneline]{subfigure}
\newcommand{\lambdaod}{\lambda^\textrm{od}}
\newcommand{\lambdad}{\lambda^\textrm{d}}

\makeatletter
\newcommand{\rmnum}[1]{\romannumeral #1}
\newcommand{\Rmnum}[1]{\expandafter\@slowromancap\romannumeral #1@}
\makeatother

\begin{document}


\title{Tunable Band Topology Reflected by Fractional Quantum Hall States\\ in Two-Dimensional Lattices}
\author{Dong Wang}
\affiliation{Institute of Physics, Chinese Academy of Sciences, Beijing 100190, China}
\author{Zhao Liu}
\email{zhaol@princeton.edu}
\affiliation{Beijing Computational Science Research Center, Beijing, 100084, China}
\affiliation{Department of Electrical Engineering, Princeton University, Princeton, New Jersey 08544, USA}
\author{Junpeng Cao}
\affiliation{Institute of Physics, Chinese Academy of Sciences, Beijing 100190, China}
\affiliation{Beijing Computational Science Research Center, Beijing, 100084, China}
\author{Heng Fan}
\email{hfan@iphy.ac.cn}
\affiliation{Institute of Physics, Chinese Academy of Sciences, Beijing 100190, China}

\date{\today}

\begin{abstract}
Two-dimensional lattice models subjected to an external effective magnetic field can form nontrivial band topologies characterized by nonzero integer band Chern numbers. In this Letter, we investigate such a lattice model originating from the Hofstadter model and demonstrate that the band topology transitions can be realized by simply introducing tunable longer-range hopping. The rich phase diagram of band Chern numbers is obtained for the simple rational flux density and a classification of phases is presented.
In the presence of interactions, the existence of fractional quantum Hall states in both $|C|=1$ and $|C|>1$ bands is
confirmed which can reflect the band topologies in different phases. In contrast, when our model reduces to a one-dimensional lattice,
the ground states are crucially different from fractional quantum Hall states. Our results may provide insights into the study of new fractional quantum Hall states and experimental realizations of various topological phases in optical lattices.
\end{abstract}

\maketitle

\emph{Introduction.}---Topologically ordered phases of matter have been attracting a great deal of interest
in modern condensed matter physics. Among various topological states, fractional quantum Hall (FQH) states provide the most prominent examples \cite{laughlin83,mr}. They support fractionally charged excitations that are essential resources for topological quantum computation \cite{topol-quantum-computing}. Although FQH states have only been observed in continuum solid state systems, it is believed that the realization of these exotic states is also feasible in lattice systems such as the optical lattice of cold atoms \cite{jz,sdl,gd,nrco,sdl,pj,pkj,mc,km,hms}.

A topological nontrivial band in the single-particle problem is an indispensable ingredient for FQH states in both the continuum and lattices \cite{tknn}.
Therefore, the investigation of band topology is usually the starting point to understand the rich FQH physics. The band topology has been studied extensively for a square lattice with complex nearest-neighbor hopping, i.e., the well-known Hofstadter model \cite{hof,mk}. Here, a question arises naturally: Can we produce novel band topology by a simple design based on the standard Hofstadter model so that new FQH physics can be obtained?

In this paper, we systematically visit this problem and discover that tunable longer-range hopping added on the conventional Hofstadter model can provide us new band topologies. We numerically calculate the band Chern numbers for simple rational flux density $\phi=p/q$, and establish various phases of band topology by tuning the strengths of nearest-neighbor and next nearest-neighbor hopping. The conventional Hofstadter model corresponds to only one phase in our rich phase diagram, and the band topologies in other phases are strikingly different from it. A classification of phases can be given for $p=1$ case. In order to identify the band topology in a many-body level, we also consider FQH states that can be harbored by topological nontrivial bands. By choosing appropriate hopping parameters, we indeed find different FQH states for different band topologies. These states are expected to be similar to recently discovered fractional Chern insulators (FCIs) \cite{chernins1,chernins2,chernins3,cherninsnum1,cherninsnum2,c1d,max,dipolar,moessner,andreas,cooper,
nonab2,qi,C2,ChernN,hcnonab,review1,review2} because of the adiabatic continuity between Hofstadter and Chern insulator states \cite{wjs,wannier,moller}.
Additionally, we study a one-dimensional lattice model for which our generalized Hofstadter model can be regarded as a two-dimensional ancestor.
However, the ground states there are crucially different from FQH states at fractional filling factors \cite{ttfci,ttfci2}.

\begin{figure}
\centerline{\includegraphics[height=3.9cm,width=0.65\linewidth] {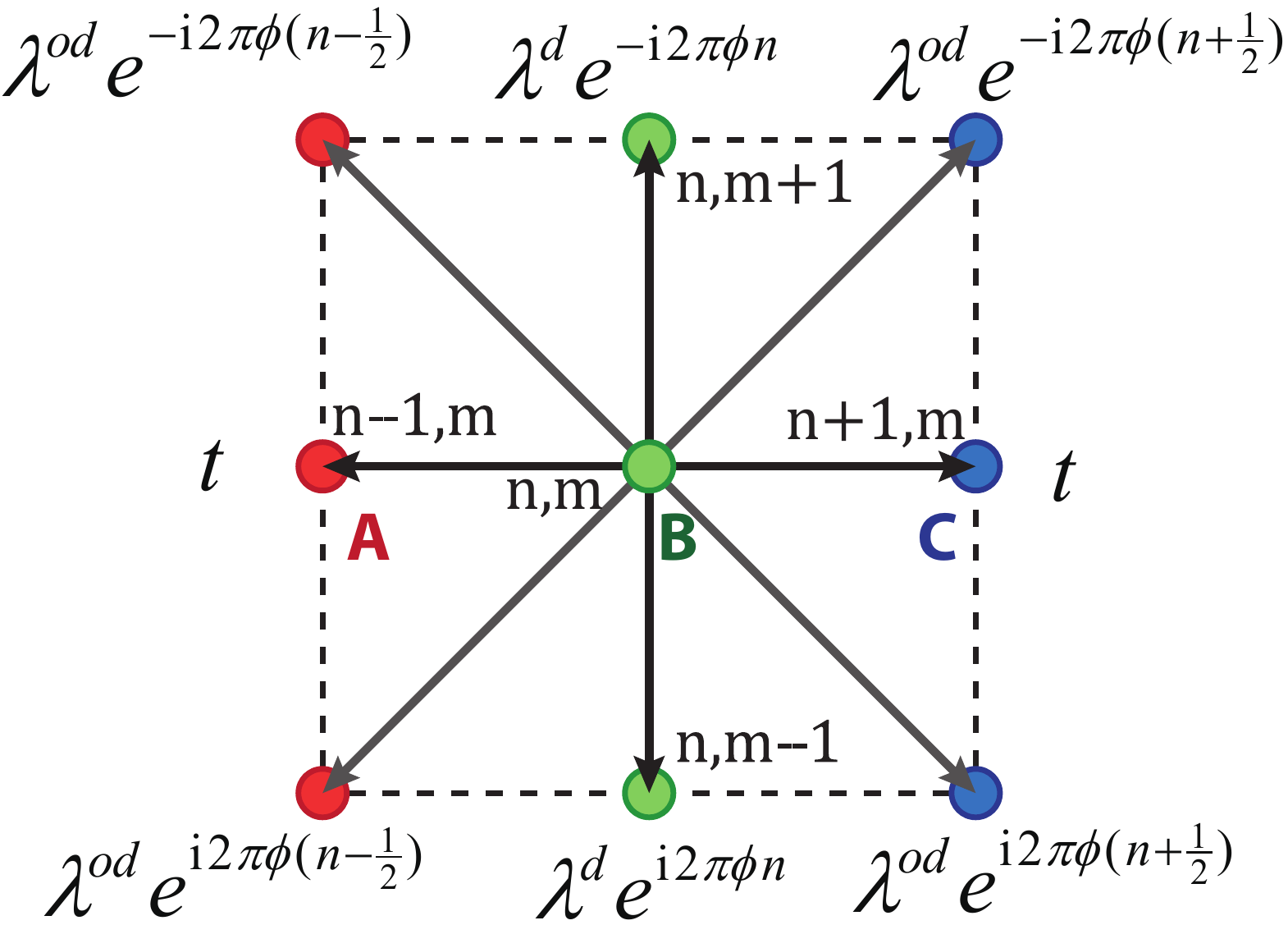}}
\caption{(Color online) An illustration of our square lattice model with $\phi=1/3$. There are three sites ($A,B,C$) per unit cell. The site coordinates and the hopping between sites are indicated explicitly. After the Fourier transform into the momentum space, the diagonal (off-diagonal) entries of the Hamiltonian only depend on $\lambdad$ ($\lambdaod$), so we use the superscripts ``d'' and ``od''.}
\label{fg:lattice}
\end{figure}

\emph{Model and band topology.}--- Here we consider a two-dimensional (2D) generalized Hofstadter model on a square lattice with tunable nearest-neighbor (NN) and next-nearest-neighbor (NNN) hopping (see Refs.~\cite{km,rachel,delgado} for other generalizations of the conventional Hofstadter model),
\begin{eqnarray}\label{eq:Hamiltonian_2D_ancestor}
H_{2\textrm{D}}=&-&\sum_{n,m}\big[t c^\dagger_{n,m}c_{n{+}1,m}{+}\lambdaod e^{i2\pi \phi (n{+}\frac12)}c^\dagger_{n,m}c_{n{+}1,m{+}1}\nonumber\\
&+&\lambdaod e^{{-}i2\pi \phi (n{+}\frac12)}c^\dagger_{n,m}c_{n{+}1,m{-}1}\nonumber\\
&+&\lambdad e^{i2\pi \phi
n}c^\dagger_{n,m}c_{n,m{+}1}+\textrm{H.c.}\big],
\end{eqnarray}
where $(n,m)$ is the site coordinate, $\phi$ is the magnetic flux through each plaquette, and $t$, $\lambdaod$ and $\lambdad$ are amplitudes of the hopping in different directions (Fig.~\ref{fg:lattice}). We set $t=1$ as the energy unit. When $\phi$ is a rational number $p/q$ ($p$ and $q$ are integers which are prime to each other), we can choose $q$ sites in the $x$ direction as a unit cell, so there are $q$ Bloch bands. Because of the complex hopping terms, the 2D lattice is pierced by a uniform perpendicular effective magnetic field which breaks the time-reversal symmetry, enabling us to label the topological property of an isolated band (without level crossings with
other bands) with eigenstate $|{\bf k}\rangle$ by its Chern number \cite{tknn}, defined as $C=\frac{1}{2\pi}\int_{\bf {k}\in \textrm{BZ}}f_{xy}({\bf k})d^2{\bf k}$, where $f_{xy}({\bf k})=\partial_x a_y-\partial_y a_x$ and
$a_j({\bf k})=-i\langle {\bf k}|\partial_j|{\bf k}\rangle,j=x,y$.

For the conventional Hofstadter model ($\lambdaod=0$), the single-particle spectrum is the well-known Hofstadter
butterfly with a fractal structure \cite{hof}, and the Chern number of each band can be described by a simple picture~\cite{mk}.
In the case of $p=1$, the middle band has $C=-(q-1)$ and other bands have $C=1$ for odd $q$, while the middle two bands have the total $C=-(q-2)$ and other bands still have $C=1$ for even $q$. In the case of $p>1$, the bands are grouped into $Q+1$ clusters by writing $p/q=1/(Q+p'/q')$, and the statement above for $p=1$ is still valid for the total Chern number in each cluster. However, in the presence of the tunable $\lambdad$ and $\lambdaod$, we do not have such a simple picture of the band topology. Instead, by numerically calculating the Chern number of each band~\cite{Fukui.numerical}, we establish a rich phase diagram of band topology.
We find that there are some critical values of $\lambdaod$ and $\lambdad$ at which the band touching occurs,
and the Chern numbers of some bands may change after crossing these critical values, indicating a phase transition of band topology.

\begin{figure}
\centerline{\includegraphics[width=\linewidth] {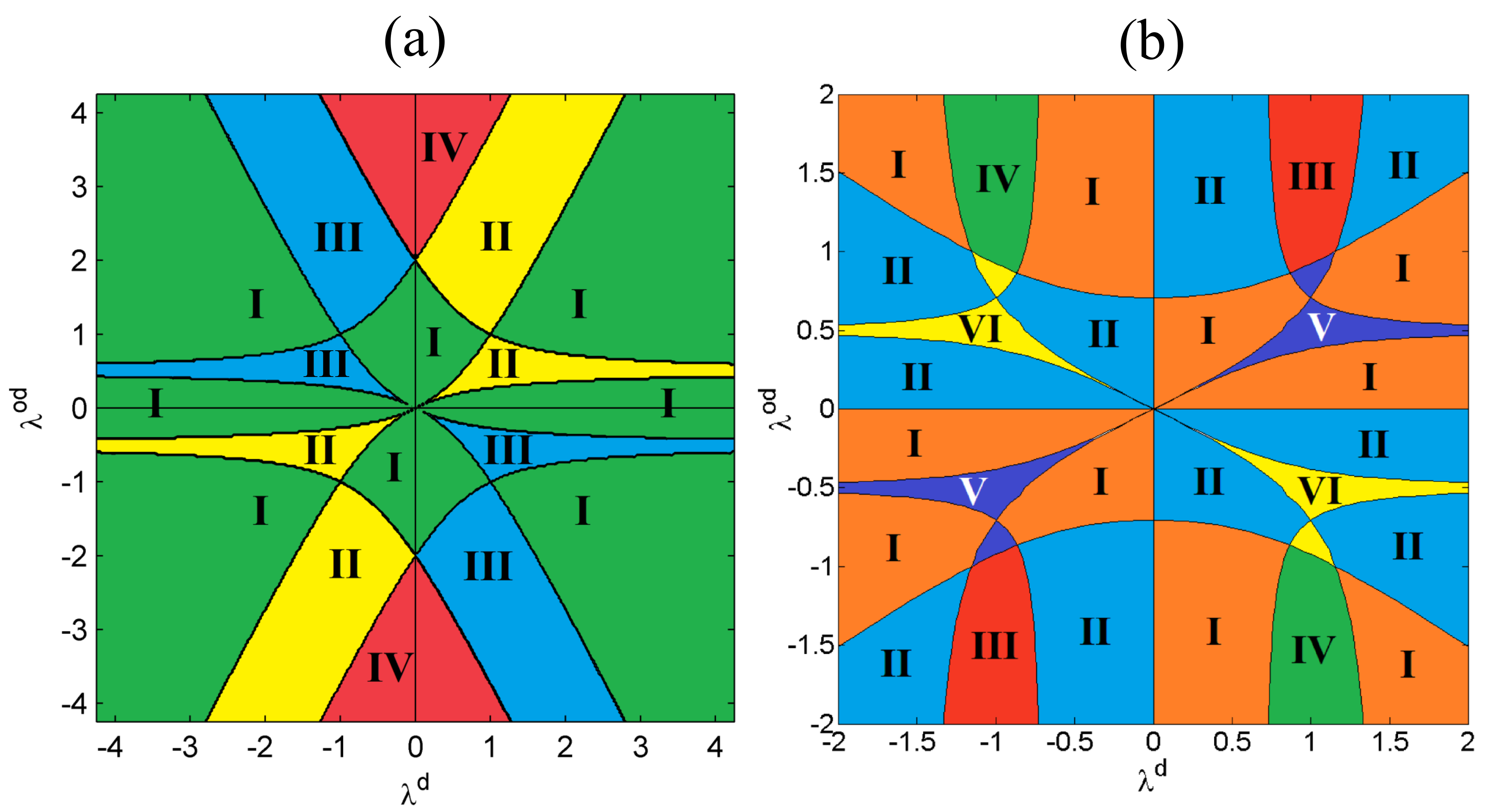}}
\caption{(Color online) The phase diagram in the $\lambdaod$--$\lambdad$ plane for (a) $\phi=1/3$ and (b) $\phi=1/4$. (a) There are four phases with different band Chern numbers: (\Rmnum{1}) $(C_1,C_2,C_3)=(1,-2,1)$; (\Rmnum{2}) $(C_1,C_2,C_3)=(1,1,-2)$; (\Rmnum{3}) $(C_1,C_2,C_3)=(-2,1,1)$; (\Rmnum{4}) $(C_1,C_2,C_3)=(-2,4,-2)$. (b) There are six phases with different band Chern numbers: (\Rmnum{1}) $(C_1,C_2,C_3,C_4)=(1,1,-3,1)$; (\Rmnum{2}) $(C_1,C_2,C_3,C_4)=(1,-3,1,1)$; (\Rmnum{3}) $(C_1,C_2,C_3,C_4)=(1,-3,5,-3)$; (\Rmnum{4}) $(C_1,C_2,C_3,C_4)=(-3,5,-3,1)$; (\Rmnum{5}) $(C_1,C_2,C_3,C_4)=(1,1,1,-3)$; (\Rmnum{6}) $(C_1,C_2,C_3,C_4)=(-3,1,1,1)$. On the boundary between two regions, band touching occurs.}
\label{fg:phase_diagram}
\end{figure}

In Fig. \ref{fg:phase_diagram}(a), we take $\phi=1/3$ as an example to demonstrate the rich phase diagram on the $\lambdad$-$\lambdaod$ plane. The whole diagram is divided into several distinct phases that are symmetric with respect to $(\lambdad,\lambdaod)=(0,0)$. In each phase,
we can label the band topology by Chern numbers of three bands from the bottom (with the lowest energy) to the top (with the highest energy), i.e. $(C_1,C_2,C_3)$ satisfying $\sum_{i=1}^q C_i=0$. The conventional Hofstadter model corresponds to the axis of $\lambdad$ (except the origin) in phase \Rmnum{1}, where $(C_1,C_2,C_3)=(1,-2,1)$ is consistent with the conclusion of Ref.~\cite{mk}. However, the band topologies in the other three phases are strikingly different from that in phase \Rmnum{1}. Chern numbers change due to the band touching and a phase transition occurs on the boundary between two neighboring phases. The sign effect of $\lambdad$ ($\lambdaod$) can be seen clearly: changing the sign of $\lambdad$ ($\lambdaod$) induces a flip of Chern numbers from $(C_1,C_2,C_3)$ to $(C_3,C_2,C_1)$. We can classify all phases into two classes according to their Chern numbers. In the first class (phase \Rmnum{1}, \Rmnum{2} and \Rmnum{3}), $(C_1,C_2,C_3)=\mathcal{P}(1,-2,1)$ ($\mathcal{P}$ means permutation), and in the second class (phase \Rmnum{4}), $(C_1,C_2,C_3)=(-2,4,-2)$.

For other values of $\phi$ with larger $q$, similar but more complicated phase diagrams can be observed. In Fig. \ref{fg:phase_diagram}(b), we show the phase diagram for $\phi=1/4$. There are six phases differentiated by Chern numbers of four bands from the bottom to the top, i.e. $(C_1,C_2,C_3,C_4)$, and the conventional Hofstadter model is located on the boundary between phase \Rmnum{1} and phase \Rmnum{2}. Similar to the $\phi=1/3$ case, the diagram is symmetric with respect to $(\lambdad,\lambdaod)=(0,0)$ and the Chern numbers are flipped by the sign change of $\lambdad$ or $\lambdaod$. All phases can also be classified into two classes. In the first class (phase \Rmnum{1}, \Rmnum{2}, \Rmnum{5} and \Rmnum{6}), $(C_1,C_2,C_3,C_4)=\mathcal{P}(1,1,-3,1)$, and in the second class (phase \Rmnum{3} and \Rmnum{4}), $(C_1,C_2,C_3,C_4)=\mathcal{P}(1,\underline{-3,5,-3})$ [the line under $(-3,5,-3)$ means that they are grouped into a cluster and should be moved as a whole in the permutation].

The symmetry and the classification of phases in the phase diagram that we observe for $\phi=1/3$ and $\phi=1/4$ are inherited by $\phi=1/5$ \cite{sm}. Therefore, we expect that there are $2q-2$ phases which can be classified into two classes (\rmnum{1}) and (\rmnum{2}) for $\phi=1/q$. We have $(C_1,C_2,...,C_{\lceil q/2 \rfloor},C_{\lceil q/2 \rfloor+1},C_{\lceil q/2 \rfloor+2},...,C_{q-1},C_q)=\mathcal{P}(1,1,...,1,1-q,1,...,1,1)$
in class (\rmnum{1}) with $q$ phases, while $(C_1,C_2,...,C_{\lceil q/2 \rfloor},C_{\lceil q/2 \rfloor+1},C_{\lceil q/2 \rfloor+2},...,C_{q-1},C_q)=\mathcal{P}(1,1,...,\underline{1{-}q,1+q,1{-q}},...,1,1)$ in class (\rmnum{2}) with $q-2$ phases, where $\lceil x \rfloor$ means the integer part of $x$. For $\phi=p/q$ with $p>1$, the situation is much more complicated. However, most of the $(\lambdad,\lambdaod)$ plane is occupied by phases in which the band Chern numbers $(C_1,...,C_q)$ can be obtained by the permutation of that for the conventional Hofstadter model \cite{sm}.

\emph{Fractional quantum Hall states.}--- In order to further characterize the band topologies of different phases discovered in the last section, we consider interacting particles partially filled in one band. If the Chern number of this occupied band changes due to a phase transition, the FQH states that this band favors to harbor should also change. Therefore, we can utilize FQH states to reflect the band topology. In the following, we choose $\phi=1/3$ to study the FQH states in various phases. These states are expected to be similar to FCIs in topological flat bands, although the net magnetic field is nonzero in our model (a gauge transformation can be used to obtain a zero net magnetic field \cite{wjs}). We adopt some commonly used criteria, such as the ground-state topological degeneracy, the spectral flow under twisted boundary conditions, and the particle-cut entanglement spectrum, to identify the ground states as FQH states \cite{cherninsnum1,cherninsnum2,c1d,max,dipolar,moessner,andreas,cooper,
nonab2,qi,C2,ChernN,hcnonab}.

We first consider $N_e$ fermions partially filled in the middle band with NN and NNN repulsive interaction
$H_{\textrm{int}}=V_{1}\sum_{\langle i,j\rangle}n_i n_j+V_2\sum_{\langle\langle i,j\rangle\rangle}n_i n_j$,
with $\langle i,j \rangle$ and $\langle\langle i,j\rangle\rangle$ representing NN and NNN sites, respectively.
In order to focus on the topological property of the band, we take the flat band limit and diagonalize
$H_{\textrm{int}}$ projected to the flattened occupied band \cite{flatexp} for a finite system on the torus with $N_1\times N_2$ unit cells. We choose three sites in the $x$ direction as a unit cell, so the actual lattice size is $3N_1\times N_2$.
The filling factor $\nu$ is defined as $N_e/(N_1N_2)$. Each energy level can be labeled by the 2D total momentum $(K_1,K_2)$.

\begin{figure}
\centerline{\includegraphics[width=\linewidth] {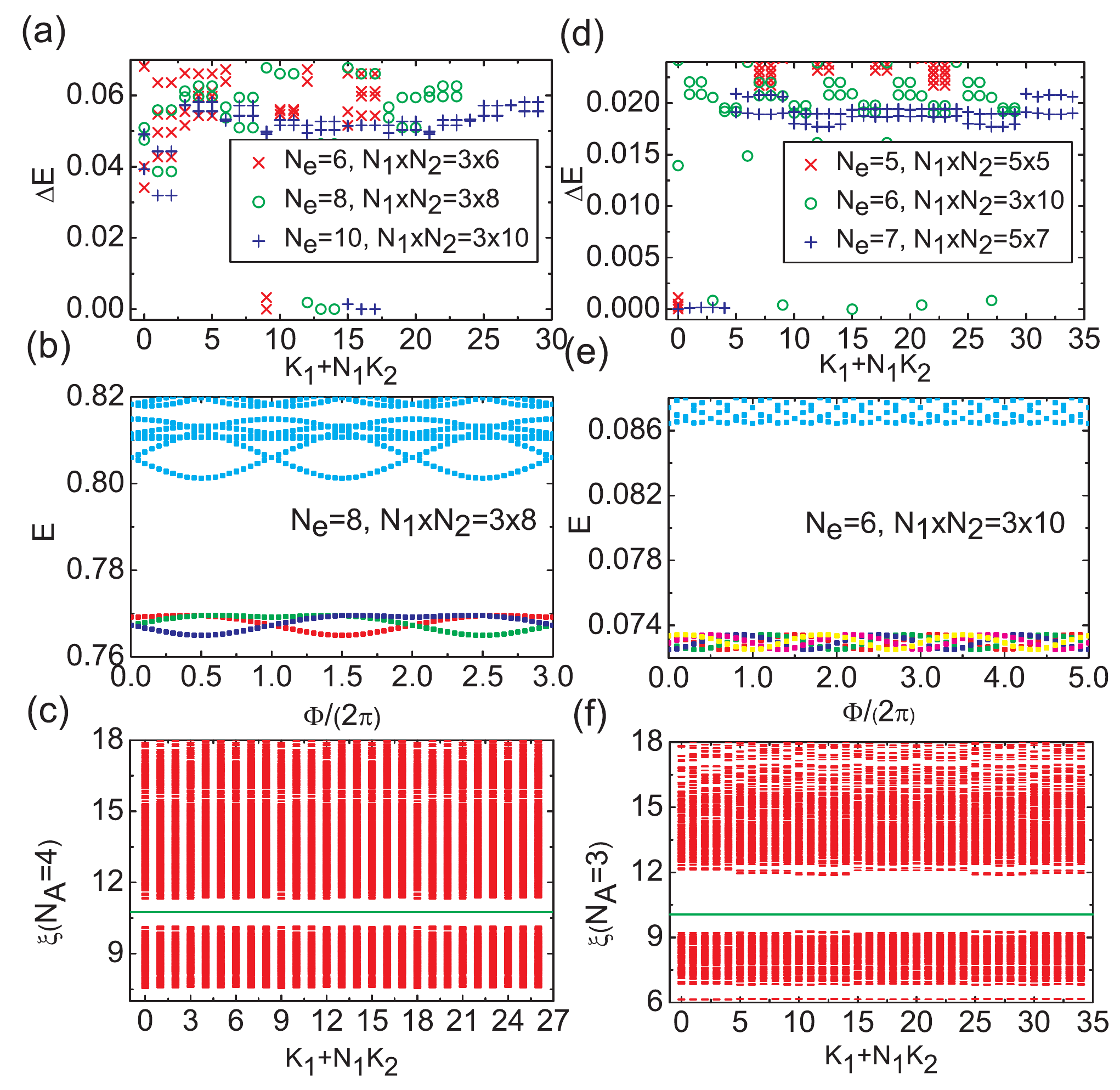}}
\caption{(Color online) Numerical results for fermionic $\nu=1/3$ [(a)-(c)] and $\nu=1/5$ [(d)-(f)] FQH states with $V_1=V_2=0.5$. $(\lambdad,\lambdaod)=(1,0.5)$ for (a)-(c) and $(\lambdad,\lambdaod)=(0.75,0)$ for (d)-(f). (a) The low-energy spectra for $N_e=6,8,10$. (b) The $x$-direction spectral flow for $N_e=8$. (c) The PES for $N_e=9$, $N_A=4$, and $N_1\times N_2=3\times9$. The number of states below the gap is $5508$. (d) The low-energy spectra for $N_e=5,6,7$. (e) The $y$-direction spectral flow for $N_e=6$. (f) The PES for $N_e=7$, $N_A=3$, and $N_1\times N_2=5\times7$. The number of states below the gap is $2695$.}
\label{fg:2D_Fermion_Harper}
\end{figure}

In phase \Rmnum{2}, the Chern number of the middle band is $C_2=1$, so the $\nu=1/3$ fermionic FQH state may be stabilized for appropriate
hopping parameters. By choosing $(\lambdad,\lambdaod)=(1,0.5)$, we indeed find three quasidegenerate ground states at the bottom of the energy spectrum separated by an energy gap from high excited levels [Fig. \ref{fg:2D_Fermion_Harper}(a)].
The spectral flow under twisted boundary conditions also confirms that the ground states are topologically nontrivial [Fig. \ref{fg:2D_Fermion_Harper}(b)]. When the boundary phase $\Phi$ changes from 0 to $3\times 2\pi$, the three ground states evolve into each other, being always separated from excited states by a gap, and finally
return to the initial configuration. In order to discard competing possibilities, such as charge density waves (CDWs), we investigate
the particle-cut entanglement spectrum (PES) \cite{cherninsnum2,ChernN,Bernevig_PRL106,Bernevig_PRB84}, which can probe the excitation structure of the ground states [Fig.~\ref{fg:2D_Fermion_Harper}(c)].
One can see a clear entanglement gap in the spectrum, and the number of low-lying levels below the gap exactly matches the quasihole counting in the corresponding Abelian FQH states. Combining all these evidences together, we are convinced that the $\nu=1/3$ fermionic FQH state
exists in the middle band in phase \Rmnum{2}.

After the system evolves from phase \Rmnum{2} to phase \Rmnum{1}, the Chern number of the middle band changes from $C_2=1$ to $C_2=-2$. Therefore,  it is expected that we can observe the $\nu=1/5$ instead of the $\nu=1/3$ fermionic FQH state for appropriate
hopping parameters \cite{ChernN}.
Our numerical results for $(\lambdad,\lambdaod)=(0.75,0)$, including the energy spectrum, the spectra flow, and the PES, provide convincing evidence
of the $\nu=1/5$ fermionic FQH state [Figs. \ref{fg:2D_Fermion_Harper}(d)-(f)] in the $C=-2$ band. Moreover, a $\nu=1/3$ state in the middle band, like that appearing in phase \Rmnum{2} is not found in phase \Rmnum{1}.
In this way, we characterize the different band topology between phase \Rmnum{1} and phase \Rmnum{2} in a many-body level.

\begin{figure}
\centerline{\includegraphics[width=\linewidth] {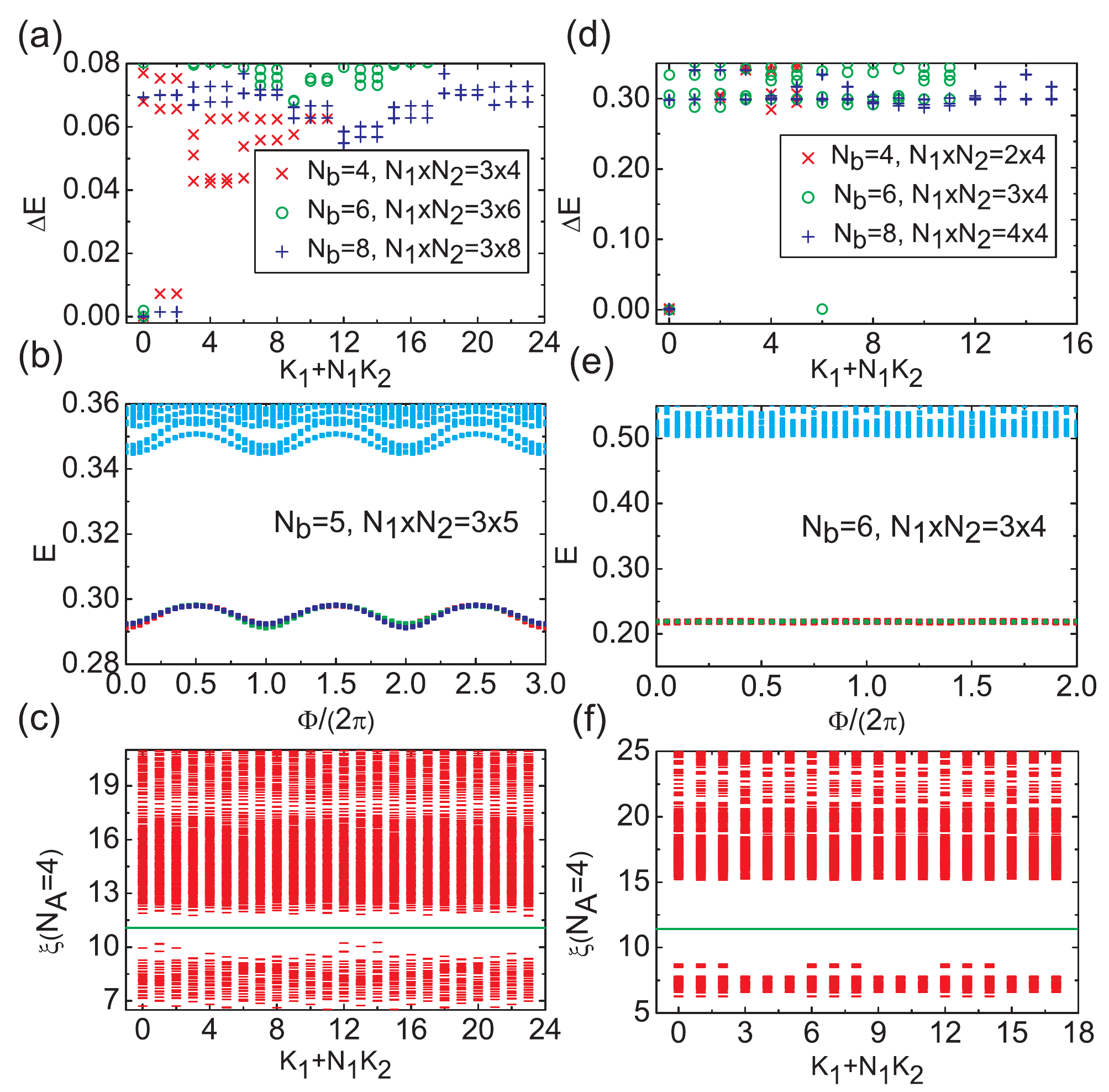}}
\caption{(Color online) Numerical results for bosonic $\nu=1/3$ [(a)-(c)] and $\nu=1/2$ [(d)-(f)] FQH states with $U=2V_1=1.0$.  $(\lambdad,\lambdaod)=(-0.75,5)$ for (a)-(c) and $(\lambdad,\lambdaod)=(1,0.5)$ for (d)-(f). (a) The low-energy spectra for $N_b=4,6,8$. (b) The $x$-direction spectral flow for $N_b=5$. (c) The PES for $N_b=8$, $N_1\times N_2=3\times8$ and $N_A=4$. The number of states below the gap is $2730$. (d) The low-energy spectra for $N_b=4,6,8$. (e) The $y$-direction spectral flow for $N_b=6$. (f) The PES for $N_b=9$, $N_1\times N_2=3\times6$ and $N_A=4$. The number of states below the gap is $1287$.}
\label{fg:2D_boson_Harper}
\end{figure}

Now we turn our attention to $N_b$ bosons partially filled in the lowest band with on-site and NN repulsive interaction
$H_{\textrm{int}}=U\sum_i n_i(n_i-1)+V\sum_{\langle i,j\rangle}n_i n_j$. Here we also flatten the occupied band and project $H_{\textrm{int}}$ to it.
In phase \Rmnum{4}, the Chern number of the lowest band is $C_1=-2$, which may harbor the unusual bosonic FQH state at $\nu=1/3$ with an odd denominator \cite{ChernN,C2}. Our numerical results indeed support the existence of this state [Figs.~\ref{fg:2D_boson_Harper}(a)-(c)].
Moreover, in phase \Rmnum{2} where $C_1$ changes to $1$ from $-2$, we find that bosons form a FQH state
at $\nu=1/2$ instead of $\nu=1/3$ [Figs.~\ref{fg:2D_boson_Harper}(d)-(f)]. Therefore, the different band topology between phase \Rmnum{2} and phase \Rmnum{4} is also confirmed in a many-body picture.

\emph{Interacting one-dimensional model.}--- Our 2D generalized Hofstadter model can be regarded as an ancestor of a one-dimensional (1D) model with the NN hopping and the on-site potential governed by a modulation of frequency $\phi$,
\begin{eqnarray}\label{eq:Hamiltonian_1D}
H_{1\textrm{D}}(\delta)=&-&\sum_n\Big\{\big[1+2\lambdaod\cos(2\pi \phi n+\delta+\pi
\phi)\big]c^\dagger_n c_{n{+}1}\nonumber\\
&&+\textrm{H.c.}+2\lambdad\cos(2\pi \phi n+\delta)c^\dagger_n c_n\Big\},
\end{eqnarray}
where $\delta$ is a phase factor that can be randomly chosen. In the case of $\lambdaod=0$, the topological properties of this model and its whole family $\{H_{1\textrm{D}}(\delta),0\leq\delta<2\pi\}$ are clarified \cite{quasicrystal}. After replacing $c_{n}$ with $c_{n,\delta}$ in Eq.~(\ref{eq:Hamiltonian_1D}), we can return to our generalized 2D Hofstadter model by $H_{2\textrm{D}}=\frac{1}{2\pi}\int_0^{2\pi}H_{1\textrm{D}}(\delta)d\delta$.

\begin{figure}
\centerline{\includegraphics[width=\linewidth] {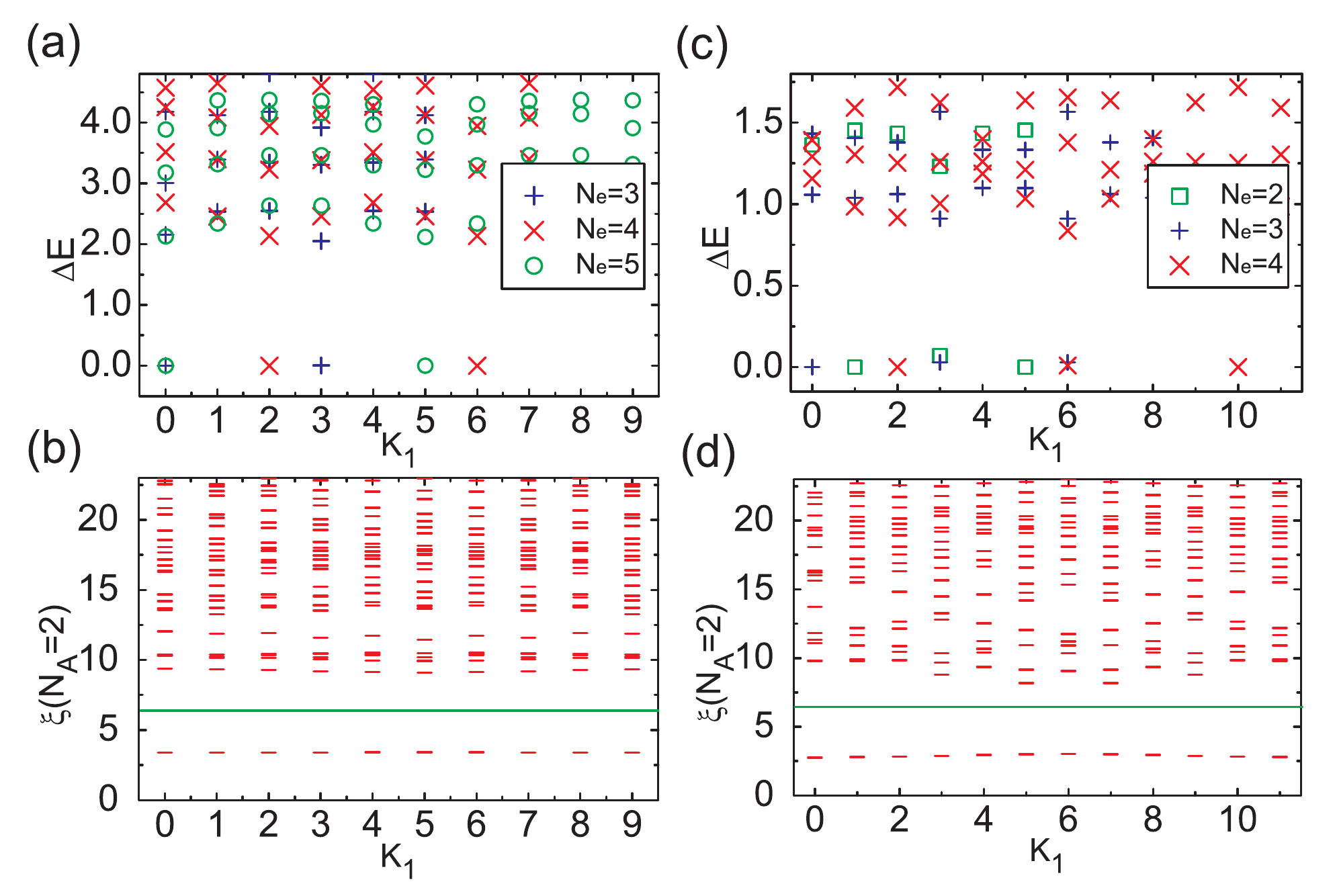}}
\caption{(Color online) Numerical results for the 1D model in Eq.~(\ref{eq:Hamiltonian_1D}) at $\nu=1/2$ [(a)-(b)] and $\nu=1/3$ [(c)-(d)], with $\delta=\pi/2$, $\phi=1/3$, and $(\lambdad,\lambdaod)=(0.75,0)$. The results are calculated without band projection in the large interaction limit. (a) The low-energy spectrum at $\nu=1/2$. (b) The PES for $N_e=5$, $N_{\textrm{cell}}=10$, and $N_A=2$. The number of states below the gap is $20$. (c) The low-energy spectrum at $\nu=1/3$. (d) The PES for $N_e=4$, $N_{\textrm{cell}}=12$, and $N_A=2$. The number of states below the gap is $18$. }\label{fg:CDW_with_DDIs}
\end{figure}

Motivated by recent discoveries of gapped phases of interacting particles in 1D flat bands at fractional fillings~\cite{ttfci2,guo,Z.H.Xu}, we consider the lattice model in Eq.~(\ref{eq:Hamiltonian_1D}) under the periodic boundary condition partially filled by $N_e$ fermions with dipole-dipole interaction $H_{\textrm{int}}=\frac{V}{2}\sum_{i\neq j}\frac{n_i n_j}{|i-j|^3}$ for $\phi=1/3$. The filling factor $\nu$ is defined as $N_e/N_{\textrm{cell}}$ with $N_{\textrm{cell}}$ the number of unit cells, each of which contains three sites. In the large interaction limit, we find that there is always an $m$-fold ground-state degeneracy at $\nu=1/m$ for $\lambdaod=0$ (Fig.~\ref{fg:CDW_with_DDIs}), even for the even $m$ where fermionic Laughlin states do not exist. Moreover, the number of states below the entanglement gap in the PES is much smaller than the FQH counting $\mathcal{N}^{N_A}_{\textrm{FQH}}=m\frac{N_e}{N_A}\left(\begin{smallmatrix}mN_e-(m-1)N_A-1\\N_A-1 \end{smallmatrix}\right)$, but matches the CDW counting $\mathcal N^{N_A}_{\textrm{CDW}}=m\left(\begin{smallmatrix}N_e\\N_A\end{smallmatrix}\right)$ \cite{ttfci,ttfci2}. Therefore, our results strongly suggest that the many-body ground states of model (\ref{eq:Hamiltonian_1D}) are CDW states in real space rather than the 1D analogues of lattice FQH states discussed in the last section.

{\it Conclusions.}--- In summary, we discover different band topologies from that in the conventional Hofstadter model by simply considering a tunable longer-range hopping. Rich phase diagrams of band topology are established on the $\lambdad$-$\lambdaod$ plane for the rational flux density $\phi=p/q$ and a classification of phases is discussed for $p=1$. The many-body FQH states that can differentiate the band topologies in various phases are also confirmed. However, the situation is completely different in 1D, where the many-body ground states have the CDW property.

There could be several future theoretical and experimental works based on our Letter, one of which may be to study the new FQH physics near rational $\phi$ \cite{hms}. Moreover, it might be interesting to study band topology transitions in other 2D and 3D lattice models \cite{other}. Very recently, a scheme of direct experimental measurement of topological invariants in optical lattices was proposed, which might be helpful to differentiate various band topologies in our phase diagrams in experiments \cite{wanglei}. Considering the hopping strength can be easily tuned in cold atom setups, our work will provide guidance for the experimental realization of various band topologies and exciting many-body fractional topological phases.

Z.~L. acknowledges E.~J.~Bergholtz, A.~M.~L\"auchli, and R.~Moessner for related collaborations and thanks E.~J.~Bergholtz for useful discussions. This work is supported by ``973'' program (2010CB922904), NSFC, and grants from CAS. Z.~L. is supported by the Department of Energy, Office of Basic Energy Sciences through Grant No.~DE-SC0002140, and the China Postdoctoral Science Foundation, Grant No. 2012M520149.

\onecolumngrid

\section*{Supplemental Material for "Tunable Band Topology Reflected by Fractional Quantum Hall States in Two-Dimensional Lattices"}
In the main text, we have shown the rich phase diagrams for $\phi=1/3$ and $\phi=1/4$. The number of distinct phases, where the nontrivial band topology is labeled by the Chern numbers, increases as the denominator of the rational flux density ($\phi=p/q$) becomes larger. Here, we will show the results for $\phi=1/5$ and $\phi=2/5$, where more phases exist than $\phi=1/3$ and $\phi=1/4$ cases. Since changing the sign of $\lambdad$ or $\lambdaod$ only flips the band Chern numbers, we only illustrate the phase diagram in the first quadrant of $\lambdad$--$\lambdaod$ plane: ($\lambdad\geq0,\lambdaod\geq0$).

\begin{figure}[h]
\centerline{\includegraphics[width=0.6\linewidth] {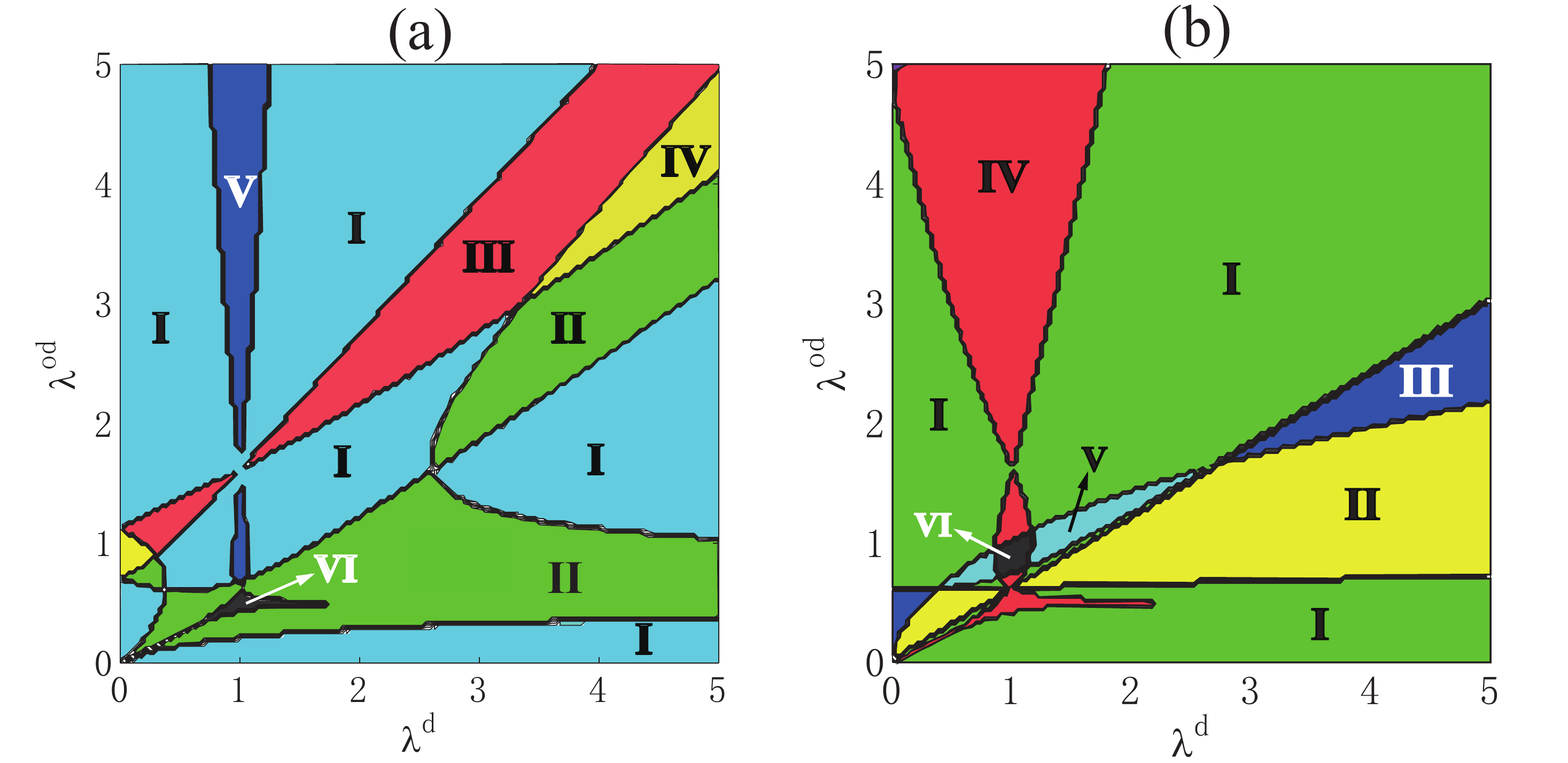}}
\caption{(Color online) (a) The phase diagram for $\phi=1/5$, with only the first quadrant ($\lambdad\geq0,\lambdaod\geq0$) being presented. Phase \Rmnum{1}-\Rmnum{6} can be seen. Phase \Rmnum{7} and \Rmnum{8} are located in the second quadrant. (b) The phase diagram for $\phi=2/5$, with only the first quadrant ($\lambdad\geq0,\lambdaod\geq0$) being presented. Phase \Rmnum{1}-\Rmnum{6} can be seen. Phase \Rmnum{7} and \Rmnum{8} are located in the second quadrant. Phase \Rmnum{9} is located beyond $\lambdaod=5$. }\label{fg:phase_diagram_sm}
\end{figure}

The phase diagram for $\phi=1/5$ is shown in Fig. \ref{fg:phase_diagram_sm}(a). Six of the eight different phases represented by different colors are illustrated in the diagram, and other two absent phases are located in the second quadrant of the $\lambdad$--$\lambdaod$ plane:  ($\lambdad<0,\lambdaod>0$). The band Chern numbers are numerically calculated in each phase, and the result is exhibited in Table \ref{tb:Chern_num_1_5th}. The conventional Hofstadter model is located in phase \Rmnum{1}. We can see that for phase \Rmnum{1} and \Rmnum{4}, the band Chern number $(C_1,C_2,C_3,C_4,C_5)$ is symmetric, i.e. $C_1=C_5$ and $C_2=C_4$. That is because, by taking the diagonal limit ($\lambdad\neq0$, $\lambdaod=0$) or the off-diagonal limit ($\lambdad=0$, $\lambdaod\neq0$), the Hamiltonian preserves the particle-hole symmetry, which makes the band Chern numbers symmetric. Furthermore, according to the data in Table \ref{tb:Chern_num_1_5th}, we can see that the eight phases can be divided into two classes: (\rmnum{1}) $(C_1,C_2,C_3,C_4,C_5)=\mathcal{P}(1,1,-4,1,1)$ and (\rmnum{2}) $(C_1,C_2,C_3,C_4,C_5)=\mathcal{P}(1,\underline{-4,6,-4},1)$, which is similar to the $\phi=1/3$ and $\phi=1/4$ cases. The classification of phases for a general $\phi=1/q$ is given in the main text.

However, the situation is different for $p{\neq}1$. Here we take $\phi=2/5$ as an example to demonstrate its different behavior from $\phi=1/5$. The phase diagram is shown in Fig. \ref{fg:phase_diagram_sm}(b), and the Chern number of each band is listed in Table \ref{tb:Chern_num_2_5th}.
The conventional Hofstadter model ($\lambdaod=0$) is located in phase \Rmnum{1}, where the band Chern numbers $(C_1,C_2,C_3,C_4,C_5)=(-2,3,-2,3,-2)$.
This is consistent with the conclusion that the five bands should be grouped into three clusters with the total Chern number $1$, $-2$ and $1$,  respectively. We write $(C_1,C_2,C_3,C_4,C_5)=(\underline{-2,3},-2,\underline{3,-2})$ to show such a clustering picture. The nine phases in Table \ref{tb:Chern_num_2_5th} can be divided into three classes: (\rmnum{1}) $(C_1,C_2,C_3,C_4,C_5)=\mathcal{P}(\underline{-2,3},-2,\underline{3,-2})$, i.e. phase \Rmnum{1}-\Rmnum{5} and \Rmnum{7}; (\rmnum{2}) $(C_1,C_2,C_3,C_4,C_5)=(-2,3,3,-7,3)$ and its inversion, i.e. phase \Rmnum{6} and \Rmnum{8}; (\rmnum{3}) $(C_1,C_2,C_3,C_4,C_5)=(-2,-2,8,-2,-2)$, i.e. phase \Rmnum{9}. One can notice that most of the phase space is occupied by phases belonging to class (\rmnum{1}).

Now, we can have a more complete look at how the conventional Hofstadter model is generalized to our model. The anisotropic hopping and NNN hopping lead to the change of band structure and band topology and more topological nontrivial phases emerge. Most of these new phases can be regarded as daughters of the conventional Hofstadter phase in the sense that their band Chern numbers are permutations of that in the father model.

\begin{table}[h]
\addtolength{\tabcolsep}{4pt}
\caption{The band Chern numbers in all phases for $\phi=1/5$. In each column, the Chern numbers are listed from band 5 on the top (with the highest energy) to band 1 at the bottom (with the lowest energy). }\label{tb:Chern_num_1_5th}
\begin{tabular}{c|rrrrrrrr}
  \hline\hline
   & \Rmnum{1} & \Rmnum{2} & \Rmnum{3} & \Rmnum{4} & \Rmnum{5} & \Rmnum{6} & \Rmnum{7} & \Rmnum{8} \\\hline
  band $5$ & $1$ & $1$ & $1$ & $1$ & $-4$ & $-4$ & $1$ & $1$ \\
  band $4$ & $1$ & $-4$ & $1$ & $-4$ & $6$ & $1$ & $1$ & $1$ \\
  band $3$ & $-4$ & $1$ & $1$ & $6$ & $-4$ & $1$ & $-4$ & $1$ \\
  band $2$ & $1$ & $1$ & $-4$ & $-4$ & $1$ & $1$ & $6$ & $1$ \\
  band $1$ & $1$ & $1$ & $1$ & $1$ & $1$ & $1$ & $-4$ & $-4$ \\
  \hline
\end{tabular}
\end{table}

\begin{table}[h]
\addtolength{\tabcolsep}{3pt}
\caption{The band Chern numbers in all phases for $\phi=2/5$. In each column, the Chern numbers are listed from band 5 on the top (with the highest energy) to band 1 at the bottom (with the lowest energy).}\label{tb:Chern_num_2_5th}
\begin{tabular}{c|rrrrrrrrr}
  \hline\hline
   & \Rmnum{1} & \Rmnum{2} & \Rmnum{3} & \Rmnum{4} & \Rmnum{5} & \Rmnum{6} & \Rmnum{7} & \Rmnum{8} & \Rmnum{9} \\\hline
  band $5$ & $-2$ & $-2$ & $ 3$ & $-2$ & $ 3$ & $ 3$ & $-2$ & $-2$ & $-2$ \\
  band $4$ & $ 3$ & $ 3$ & $-2$ & $-2$ & $-2$ & $-7$ & $ 3$ & $ 3$ & $-2$ \\
  band $3$ & $-2$ & $-2$ & $-2$ & $ 3$ & $-2$ & $ 3$ & $ 3$ & $ 3$ & $ 8$ \\
  band $2$ & $ 3$ & $-2$ & $-2$ & $ 3$ & $ 3$ & $ 3$ & $-2$ & $-7$ & $-2$ \\
  band $1$ & $-2$ & $ 3$ & $ 3$ & $-2$ & $-2$ & $-2$ & $-2$ & $ 3$ & $-2$ \\
  \hline
\end{tabular}
\end{table}


\begin{thebibliography}{99}

\bibitem{laughlin83}
R.~B.~Laughlin,  Phys.~Rev.~Lett.~{\bf 50}, 1395 (1983).

\bibitem{mr}
G.~Moore, and N.~Read, Nucl.~Phys.~B {\bf 360}, 362 (1991).

\bibitem{topol-quantum-computing}
C.~Nayak, S.~H.~Simon, A.~Stern, M.~Freedman, and S.~Das Sarma,
Rev.\ Mod.\ Phys.\ {\bf 80}, 1083 (2008).

\bibitem{jz}
D. Jaksch and P. Zoller, New J. Phys. {\bf 5}, 56 (2003).

\bibitem{sdl}
A. S. S{\o}rensen, E. Demler, and M. D. Lukin, Phys. Rev. Lett. {\bf 94}, 086803 (2005).

\bibitem{gd}
F. Gerbier and J. Dalibard, New J. Phys. {\bf 12}, 033007 (2010).

\bibitem{nrco}
N. R. Cooper, Phys. Rev. Lett. {\bf 106}, 175301 (2011).

\bibitem{pj}
R. N. Palmer and D. Jaksch, Phys. Rev. Lett. {\bf 96}, 180407 (2006).

\bibitem{pkj}
R. N. Palmer, A. Klein, and D. Jaksch, Phys.~Rev.~A~{\bf 78}, 013609~(2008).

\bibitem{mc}
G. M\"{o}ller and N. R. Cooper, Phys. Rev. Lett. {\bf 103}, 105303 (2009).

\bibitem{km}
E. Kapit and E. Mueller, Phys. Rev. Lett. {\bf 105}, 215303 (2010).

\bibitem{hms}
L. Hormozi, G. M\"{o}ller, and S. H. Simon, Phys. Rev. Lett. {\bf 108}, 256809 (2012).

\bibitem{tknn}
D. J. Thouless, M. Kohmoto, M. P. Nightingale, and M. den Nijs, Phys. Rev. Lett. {\bf 49}, 405 (1982).

\bibitem{hof}
D. R. Hofstadter, Phys. Rev. B {\bf 14}, 2239 (1976).

\bibitem{mk}
M. Kohmoto, Phys. Rev. B {\bf 39}, 11943 (1989).

\bibitem{chernins1}
E.~Tang, J.-W.~Mei, and X.-G.~Wen, Phys.~Rev.~Lett.~{\bf 106}, 236802 (2011).

\bibitem{chernins2}
K.~Sun, Z.~Gu, H.~Katsura, and S.~Das Sarma, Phys.~Rev.~Lett.~{\bf 106}, 236803 (2011).

\bibitem{chernins3}
T.~Neupert, L.~Santos, C.~Chamon, and C.~Mudry, Phys.~Rev.~Lett.~{\bf 106}, 236804 (2011).

\bibitem{cherninsnum1}
D.~N.~Sheng, Z.~Gu, K.~Sun, and L.~Sheng, Nat.~Commun.~{\bf 2}, 389 (2011).

\bibitem{cherninsnum2}
N.~Regnault and B.~A.~Bernevig,  Phys. Rev. X {\bf 1}, 021014 (2011); B.~A.~Bernevig and N.~Regnault, Phys. Rev. B {\bf 85}, 075128 (2012); Y.-L.~Wu, B.~A.~Bernevig, and N.~Regnault, Phys. Rev. B {\bf 85}, 075116 (2012); T.~Liu, C.~Repellin, B.~A.~Bernevig, and N.~Regnault, Phys. Rev. B {\bf 87}, 205136 (2013); Y.-L.~Wu, N.~Regnault, and B.~A.~Bernevig, Phys. Rev. B {\bf 86}, 085129 (2012).

\bibitem{c1d}
J.~W.~F.~Venderbos, S.~Kourtis, J.~van den Brink, and M.~Daghofer, Phys. Rev. Lett. {\bf 108}, 126405 (2012).

\bibitem{dipolar}
N.~Y.~Yao,  A.~V.~Gorshkov, C.~R.~Laumann, A.~M.~L\"auchli, J.~Ye, and M.~D.~Lukin, Phys. Rev. Lett. {\bf 110}, 185302 (2013).

\bibitem{moessner}
N.~R.~Cooper and R.~Moessner, Phys. Rev. Lett. \textbf{109}, 215302 (2012).

\bibitem{andreas}
A.~M.~L\"auchli, Z.~Liu, E.~J.~Bergholtz, and R.~Moessner, Phys.~Rev.~Lett. {\bf 111}, 126802 (2013).

\bibitem{cooper}
N.~R.~Cooper and J.~Dalibard, Phys. Rev. Lett. {\bf 110}, 185301 (2013).

\bibitem{nonab2}
Y.-F. Wang, Z.-C. Gu, C.-D. Gong, and D. N. Sheng,
Phys. Rev. Lett. {\bf 107}, 146803 (2011); Y.-F.~Wang, H.~Yao, Z.-C.~Gu, C.-D.~Gong, and D.~N.~Sheng,  Phys. Rev. Lett. {\bf 108}, 126805 (2012).

\bibitem{qi}
X.-L.~Qi, Phys. Rev. Lett. {\bf 107}, 126803 (2011).

\bibitem{max}
M.~Trescher and E.~J.~Bergholtz, Phys. Rev. B {\bf 86}, 241111(R) (2012).

\bibitem{C2}
Y.-F. Wang, H. Yao, C.-D. Gong, and D. N. Sheng, Phys. Rev. B {\bf 86}, 201101(R) (2012).

\bibitem{ChernN}
Z.~Liu, E.~J.~Bergholtz, H.~Fan, and A.~M.~L\"auchli, Phys. Rev. Lett. {\bf 109}, 186805 (2012).

\bibitem{hcnonab}
A.~Sterdyniak, C.~Repellin, B.~A.~Bernevig, and N.~Regnault, Phys. Rev. B {\bf 87}, 205137 (2013);
Y.-L. Wu, N. Regnault, and B. A. Bernevig, Phys. Rev. Lett. {\bf 110}, 106802 (2013).

\bibitem{review1}
S. A. Parameswaran, R. Roy, and S. L. Sondhi, arXiv: 1302.6606.

\bibitem{review2}
E. J. Bergholtz and Z. Liu, Int. J. Mod. Phys. B {\bf 27}, 1330017 (2013).

\bibitem{wjs}
Y.-H. Wu, J. K. Jain, and K. Sun, Phys. Rev. B {\bf 86}, 165129 (2012).

\bibitem{wannier}Z.~Liu and E.~J.~Bergholtz, Phys. Rev. B {\bf 87}, 035306 (2013).

\bibitem{moller}T.~Scaffidi, and G.~M\"oller,
Phys. Rev. Lett. {\bf 109}, 246805 (2012).

\bibitem{ttfci}
B. A. Bernevig and N. Regnault, arXiv: 1204.5682.

\bibitem{ttfci2}
J. C. Budich and E. Ardonne, Phys. Rev. B {\bf 88}, 035139 (2013).

\bibitem{rachel}
D. Cocks, Peter P. Orth, S. Rachel, M. Buchhold, K. Le Hur, and W. Hofstetter, Phys. Rev. Lett. {\bf 109}, 205303 (2012).

\bibitem{delgado}
N. Goldman, A. Kubasiak, A. Bermudez, P. Gaspard, M. Lewenstein, and M. A. Martin-Delgado, Phys. Rev. Lett. {\bf 103}, 035301 (2009);
N. Goldman, I. Satija, P. Nikolic, A. Bermudez, M. A. Martin-Delgado, M. Lewenstein, and I. B. Spielman Phys. Rev. Lett. {\bf 105}, 255302 (2010).

\bibitem{Fukui.numerical}
T. Fukui, Y. Hatsugai, and H. Suzuki, J. Phys. Soc. Jpn. {\bf 74}, 1674 (2005).

\bibitem{sm}
See the Supplemental Material for the phase diagrams for $\phi=1/5$ and $\phi=2/5$.

\bibitem{flatexp}
This means that, we do not consider the dispersion of the occupied band or the scattering of particles between different bands. This strategy can isolate the effect of interactions, significantly reduce the dimension of the many-body Hilbert space and increase the numerical efficiency (see, e.g., Ref.~\cite{cherninsnum2} for details).

\bibitem{Bernevig_PRL106}
A. Sterdyniak, N. Regnault, and B.A. Bernevig, Phys. Rev. Lett. {\bf 106}, 100405 (2011).

\bibitem{Bernevig_PRB84}
A. Chandran, M. Hermanns, N. Regnault, and B.A. Bernevig, Phys. Rev. B {\bf 84}, 205136 (2011).

\bibitem{quasicrystal}
K. A. Madsen, E. J. Bergholtz, and P. W. Brouwer, Phys.~Rev.~B {\bf 88}, 125118 (2013).

\bibitem{guo}
H.-M. Guo, S.-Q. Shen, and S.-P. Feng, Phys. Rev. B {\bf 86}, 085124 (2012); H.-M. Guo, Phys. Rev. A {\bf 86}, 055604 (2012).

\bibitem{Z.H.Xu}
Z.-H. Xu, L.-H. Li, and S. Chen, Phys. Rev. Lett. {\bf 110}, 215301 (2013).

\bibitem{other}
A. Bermudez, L. Mazza, M. Rizzi, N. Goldman, M. Lewenstein, and M. A. Martin-Delgado, Phys. Rev. Lett. {\bf 105}, 190404 (2010);
A. Bermudez, N. Goldman, A. Kubasiak, M. Lewenstein, and M. A. Martin-Delgado, New J. Phys. {\bf 12}, 033041 (2010).

\bibitem{wanglei}
L. Wang, A. A. Soluyanov, and M. Troyer, Phys. Rev. Lett. {\bf 110}, 166802 (2013).

\end{thebibliography}
\end{document}